\documentclass[letterpaper]{article} % DO NOT CHANGE THIS
\usepackage{aaai2026}  % DO NOT CHANGE THIS
\usepackage{times}  % DO NOT CHANGE THIS
\usepackage{helvet}  % DO NOT CHANGE THIS
\usepackage{courier}  % DO NOT CHANGE THIS
\usepackage[hyphens]{url}  % DO NOT CHANGE THIS
\usepackage{graphicx} % DO NOT CHANGE THIS
\usepackage{natbib}  % DO NOT CHANGE THIS AND DO NOT ADD ANY OPTIONS TO IT
\usepackage{caption} % DO NOT CHANGE THIS AND DO NOT ADD ANY OPTIONS TO IT
\usepackage[linesnumbered,ruled,vlined]{algorithm2e}

\usepackage{amsmath,amssymb,amsfonts} % math

\usepackage{multirow}
\usepackage{booktabs}
\usepackage{tabularx}
\usepackage{threeparttable}

\usepackage{newfloat}
\usepackage{listings}
\DeclareCaptionStyle{ruled}{labelfont=normalfont,labelsep=colon,strut=off} % DO NOT CHANGE THIS
\newcommand{\dataset}{ForgeHLS}
\newcommand{\kernelnum}{846}
\newcommand{\designnum}{459,850}

\usepackage{pifont} %生成图表中的对勾和叉号
\newcommand{\no}{\ding{56}}   % 叉（黑色）
\newcommand{\ye}{\ding{52}}   % 勾（黑色） % 叉（黑色）
\newcommand{\incomplete}{\ding{115}} %黑色三角

\nocopyright

\title{\dataset{}: A Large-Scale, Open-Source Dataset for High-Level Synthesis}
\author{
    Zedong Peng\textsuperscript{\rm 1}, 
    Zeju Li\textsuperscript{\rm 2}, 
    Mingzhe Gao\textsuperscript{\rm 1}, 
    Qiang Xu\textsuperscript{\rm 2}, 
    Chen Zhang\textsuperscript{\rm 1}, 
    Jieru Zhao\textsuperscript{\rm 1 }
}

\affiliations{
    \textsuperscript{\rm 1}Shanghai Jiao Tong University, Shanghai, China\\
    \textsuperscript{\rm 2}The Chinese University of Hong Kong, Hong Kong S.A.R.\\
    \{zedongpeng, zhao-jieru\}@sjtu.edu.cn
}

\begin{document}

\maketitle

\begin{abstract}
High-Level Synthesis (HLS) plays a crucial role in modern hardware design by transforming high-level code into optimized hardware implementations. However, progress in applying machine learning (ML) to HLS optimization has been hindered by a shortage of sufficiently large and diverse datasets. To bridge this gap, we introduce ForgeHLS, a large-scale, open-source dataset explicitly designed for ML-driven HLS research. ForgeHLS comprises over 400k diverse designs generated from \kernelnum{} kernels covering a broad range of application domains, consuming over 200k CPU hours during dataset construction. Each kernel includes systematically automated pragma insertions (loop unrolling, pipelining, array partitioning), combined with extensive design space exploration using Bayesian optimization. Compared to existing datasets, ForgeHLS significantly enhances scale, diversity, and design coverage. We further define and evaluate representative downstream tasks in Quality of Result (QoR) prediction and automated pragma exploration, clearly demonstrating ForgeHLS utility for developing and improving ML-based HLS optimization methodologies. The dataset and code are
public at https://github.com/zedong-peng/ForgeHLS.

\end{abstract}

%     \link{Code}{https://github.com/}
%     \link{Datasets}{https://huggingface.co/datasets/}

\section{Introduction}

High-Level Synthesis (HLS) is a key technology in modern hardware design, enabling the transformation of high-level code, such as C, C++, or SystemC, into optimized hardware implementations, such as Verilog and VHDL. HLS serves as a bridge between software and hardware design, offering designers the ability to automatically generate hardware descriptions from high-level algorithmic specifications. This process significantly reduces the time and complexity traditionally required in hardware design, making it a pivotal tool for accelerating the development of custom hardware solutions. A critical aspect of HLS is the use of pragmas. Pragmas are directives that guide hardware optimizations without altering algorithmic code, influencing performance, resource utilization, and power consumption.
% There are two ways to insert pragmas: by manually adding directive files, and by inserting pragmas directly in the source code.

\begin{table}[t!]
	\centering
    \begin{threeparttable}

    \footnotesize
	\begin{tabular}{l@{\hspace{0.5em}}l@{\hspace{-0.45em}}l@{\hspace{-0.05em}}l@{\hspace{-1.35em}}l@{\hspace{-1.5em}}l}
		\toprule
		\textbf{Feature}              & \rotatebox{45}{\textbf{DB4HLS}} & \rotatebox{45}{\textbf{HLSyn}} & \rotatebox{45}{\textbf{HLSDataset}} & \rotatebox{45}{\textbf{HLSFactory}} & \rotatebox{45}{\textbf{ForgeHLS}} \\

		\hline
		\textbf{Source}                \\
		CHStone                       & \no             & \no             & \incomplete         & \incomplete         & \ye               \\
		MachSuite                     & \ye             & \ye             & \incomplete                 & \ye                 & \ye               \\
		Polybench                     & \no             & \ye             & \ye         & \incomplete         & \ye               \\
		Rosetta                       & \no             & \no             & \incomplete                 & \ye                 & \ye               \\
		Vitis-Examples                & \no             & \no             & \no         & \incomplete         & \ye               \\
        Vitis-Library                & \no             & \no             & \no         & \no         & \ye               \\
		Synthetic Code                & \no             & \no             & \no                 & \no                 & \ye               \\
		\midrule
        \textbf{Pre-HLS} \\
		Code w/ Pragma            & \ye             & \incomplete             & \no                 & \no                 & \ye               \\
        Unified File Structure            & \ye             & \ye             & \no                 & \no                 & \ye               \\

        \midrule
		\textbf{Post-HLS} \\
		Latency                       & \ye             & \ye             & \incomplete                 & \incomplete                 & \ye               \\
		Resources                     & \ye             & \ye             & \incomplete                 & \incomplete                 & \ye               \\
		Graph                 & \no             & \incomplete             & \incomplete                 & \no                 & \ye               \\
        Verilog Code & \no             & \no             & \ye                 & \ye                 & \ye               \\
		\bottomrule
	\end{tabular}

    \caption{Feature Comparison of HLS Datasets. \textbf{Code w/ Pragma}: the source code of HLS designs directly inserted with pragmas, rather than defined in separate file; \textbf{Graph}: post-HLS files used to generate graph;  \ye: fully supported; \no: not supported; \incomplete: partially supported, to be released, or requiring additional operations (e.g., re-running synthesis on the provided HLS code) to access.}
	\label{feature_comparison}
    \end{threeparttable}    
\end{table}

HLS research has focused on several key downstream tasks, including QoR (Quality of Result) prediction~\cite{h-gnn, wu2022high, lin2022powergear,9045442}, and DSE (Design Space Exploration)
\cite{sohrabizadeh2021autodseenablingsoftwareprogrammers, 10416120}. QoR prediction aims to estimate the latency and resource utilization from high-level code. DSE focuses on systematically exploring different HLS pragma configurations to identify optimal design points that balance performance, resource usage and power consumption.
% , making pragma selection a central challenge in HLS optimization. 
Both tasks rely on large and high-quality datasets to improve the efficiency and effectiveness of HLS tools. However, existing datasets for HLS~\cite{Db4hls,abi2024hlsfactory,wei2023hlsdataset,bai2023hlsyn}  have been limited in scope and feature, shown in Table~\ref{tab:statistics_comparison}.

\begin{figure*}[t!]
    \centering
    \includegraphics[width=\linewidth]{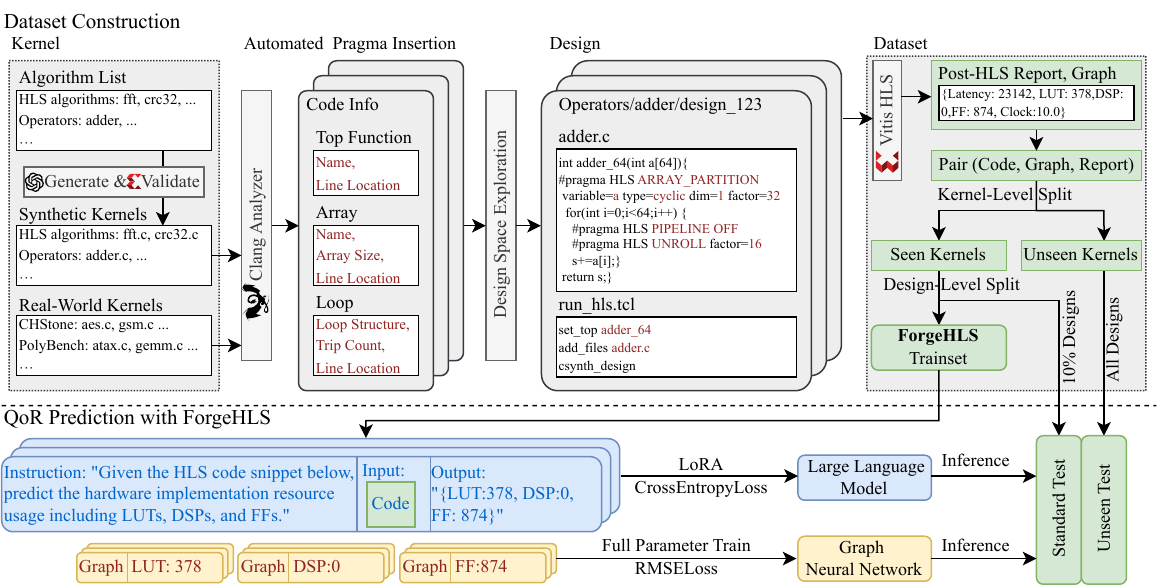}
    \caption{The overview workflow of our dataset construction and evaluation.}
    \label{fig:overview}
\end{figure*}

To address these challenges, we introduce \textbf{ForgeHLS}, a large-scale, open-source dataset featuring \designnum{} HLS designs generated from \kernelnum{} diverse C++ kernels. ForgeHLS is more than 10× larger than existing HLS datasets and spans a broad spectrum of algorithms across various domains. Figure~\ref{fig:overview} illustrates our comprehensive data construction and evaluation methodology.

\textbf{Diverse Algorithm Coverage.} \dataset{} collects real-world HLS kernels from established benchmarks as well as official AMD Xilinx collections~\cite{Vitis_Examples, Vitis_Libraries}, and unifies them into a consistent format. Moreover, we introduce synthetic code to enhance our dataset. We collect a diverse list of algorithms, leverage GPT-4o to generate corresponding HLS kernel implementations and perform validation to ensure the quality of the synthetic code.

\textbf{Automated Pragma Insertion.} We developed an automated code analysis workflow that extracts key information from HLS kernel code and explore the full design space by generating all possible combinations of pragma types and factors. For large-scale kernels that cannot be fully explored through exhaustive enumeration, we utilize Bayesian optimization to guide the design space exploration. 
% In previous datasets, the combinations of pragmas are typically defined in a script or a separate file~\cite{wei2023hlsdataset,abi2024hlsfactory, bai2023hlsyn}, requiring users to generate a large amount of feasible HLS designs for training and testing. This process incurs additional substantial effort and time cost.
% are typically defined separately using directives~\cite{wei2023hlsdataset} or domain-specific languages~\cite{abi2024hlsfactory}, 
% which require additional substantial effort and time to generate a large amount of feasible HLS designs with pragmas inserted.
% additional configuration in the TCL file or the use of external source code. 
% In contrast, we provide feasible HLS designs with corresponding post-HLS performance metrics, and directly insert pragmas into source code to make generated HLS designs eaiser to use.
% By using a simple HLS TCL file, we can run HLS with the inserted pragmas. 
% Our direct pragma insertion allows the code to be easily used for large language model (LLM) and other downstream tasks. This automated pragma insertion workflow will also be open-source, ensuring the dataset's extensibility to support additional pragma types in the future. 
By inserting comprehensive pragmas, we obtained \designnum{} HLS designs.

\begin{table}[t!]
    \centering
    \begin{threeparttable}

    \footnotesize
    \begin{tabular}{l@{\hspace{1em}}c@{\hspace{0.8em}}c@{\hspace{0.8em}}c@{\hspace{0.8em}}c@{\hspace{0.8em}}c}
        \toprule
         & \textbf{\#K} & \textbf{\#D} & \textbf{Avg. \#P} & \textbf{Avg. \#T} & \textbf{\#S} \\ 
        \midrule
        
        \multicolumn{6}{l}{\textbf{Related Work}} \\
        DB4HLS & 19 & 124,106 & 6.5*& 2270.7* & 0.1* \\
        HLSyn & 42 & 42,000 & 7.6 & 629.9 & 0.1* \\
        HLSDataset & 34 & 18,876 & 22.9* & 1144.5* & 150* \\
        HLSFactory & 82* & 3,206* & 16.4* & 520.7* & 2.5* \\
        \midrule
        
        \multicolumn{6}{l}{\textbf{Ours in Summary}} \\
        R-ForgeHLS & 405 & 99,601 & 192.4 & 1298.2 & 1169 \\
        S-ForgeHLS & 441 & 360,249 & 4.9 & 165.8 & 1008 \\
        \textbf{ForgeHLS} & \textbf{\kernelnum{}} & \textbf{\designnum{}} & \textbf{95.3} & \textbf{718.6} & \textbf{2177} \\
        \midrule

        \multicolumn{6}{l}{\textbf{ForgeHLS Subset}} \\
        \multicolumn{6}{l}{\textbf{R-ForgeHLS}} \\
        CHStone & 4 & 4,684 & 24.0 & 5932.7 & 40 \\
        MachSuite & 19 & 16,803 & 18.9 & 990.8 & 253 \\
        PolyBench & 26 & 22,897 & 18.6 & 486.2 & 412 \\
        Rosetta & 5 & 1,849 & 197.8 & 8985.8 & 115 \\
        Vitis-Examples & 41 & 22,794 & 8.2 & 673.8 & 65 \\
        Vitis-Library & 310 & 30,574 & 244.0  & 1365.9 & 284 \\
        \multicolumn{6}{l}{\textbf{S-ForgeHLS}} \\
        HLS algorithms & 113 & 93,379 & 8.5 & 285.6 & 492 \\
        Leetcode & 22 & 18,780 & 5.6 & 203.2 & 47 \\
        Operators & 154 & 85,099 & 5.5 & 165.7 & 150 \\
        RTL chip & 16 & 12,922 & 6.9 & 199.8 & 76 \\
        RTL ip & 5 & 4,874 & 5.6 & 279.5 & 66 \\
        RTL module & 131 & 145,195 & 5.2 & 128.2 & 177 \\

        \bottomrule
    \end{tabular}

    \caption{Statistics Comparison of HLS Datasets. \textbf{\#K}: number of kernels; \textbf{\#D}: number of designs; \textbf{Avg. \#P}: average number of pragmas per kernel; \textbf{Avg. \#T}: average number of tokens per kernel by \texttt{cl100k\_base} encoder (same as GPT-4); \textbf{\#S}: total size of the folder in GB; \textbf{R-ForgeHLS}: real-world part of ForgeHLS; \textbf{S-ForgeHLS}: synthetic part of ForgeHLS; \textbf{*}: calculated by us.}
    \label{tab:statistics_comparison}

    % \caption{Statistics Comparison of HLS Datasets.}
    % \label{tab:statistics_comparison}
    % \begin{tablenotes}
    %     \footnotesize
    %     \item \textbf{\#K}: total number of kernels; \textbf{\#D}: total number of designs; \textbf{Avg. \#P}: average number of pragmas per kernel; \textbf{Avg. \#T}: average number of tokens per kernel, calculated by tokenizing all C/C++ source files for each kernel using the \texttt{cl100k\_base} encoder (same as GPT-4); \textbf{\#S}: total size of the folder in gigabytes (GB); \textbf{R-ForgeHLS}:  real-world part of ForgeHLS; \textbf{S-ForgeHLS}: synthetic part of ForgeHLS; \textbf{-}: unavailable or not reported; *:  Data calculated by us.
    % \end{tablenotes}

    \end{threeparttable}
\end{table}

\textbf{Multi-task Evaluation.} 
After constructing our comprehensive dataset, a series of HLS downstream tasks can be investigated and conducted.
Specifically, we present two representative tasks: HLS QoR (Quality of results) prediction and HLS Pragma Exploration.

In summary, our contributions offer a valuable foundation for the ML community, promoting advancements in model performance and accelerating machine learning-driven innovation in HLS.
Our main contributions are:
\begin{enumerate}
    \item We propose a large-scale, diverse HLS dataset that covers a broad spectrum of algorithms and application domains.
	\item We propose an automated workflow for pragma insertion, directly embedding HLS pragmas into our kernel code.
    \item We define two representative HLS downstream tasks to evaluate HLS dataset.  
\end{enumerate}

\section{Related Work}

\subsection{HLS Datasets}
\label{sec:related-dataset}

\begin{figure*}[t]
    \centering
    \includegraphics[width=\linewidth]{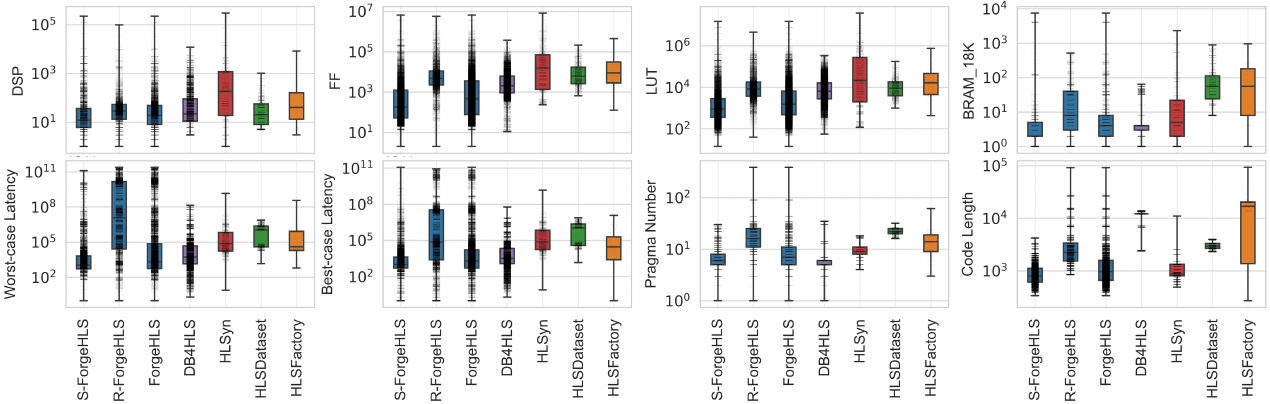}
\caption{Box plot distribution comparison of post-HLS feature across HLS datasets. Metrics include LUT (Look-Up Table), DSP (Digital Signal Processor), FF (Flip-Flop), BRAM\_18K (18-Kbit Block RAM), worst-case latency in cycles, best-case latency in cycles, pragma number applied per design and code length (measured by character count). Each dot represents an individual design. ForgeHLS exhibits significantly denser coverage across the design space; in some dimensions, the data points are so concentrated that they form visually continuous bands. Each box shows the spread of the middle 50\% of data.}
% (from Q1 to Q3).}

	\label{fig:distribution_comparison}
\end{figure*}

Numerous efforts have been made to develop datasets and frameworks for HLS research, as listed in Table \ref{feature_comparison}.
There are some classic small-scale HLS benchmarks that are commonly favored in mainstream training, including PolyBench~\cite{Polybench}, CHStone~\cite{hara2008chstone}, MachSuite~\cite{reagen2014machsuite}, and Rosetta~\cite{zhou2018rosetta}. These benchmarks provide kernel source code, and many HLS datasets are derived from these kernels through DSE to construct their datasets.
These hand-crafted kernels are well-structured and suitable for algorithmic analysis and compiler optimization tasks. However, they cover only a narrow range of application domains.
For large-scale HLS dataset, DB4HLS~\cite{Db4hls} provided a database of over 100,000 HLS design points collected from MachSuite through exhaustive design space exploration. HLSyn~\cite{bai2023hlsyn} offers 42,000 labeled designs from MachSuite and PolyBench. HLSDataset~\cite{wei2023hlsdataset} and HLSFactory~\cite{abi2024hlsfactory} extended their datasets on more benchmarks. 
However, current HLS datasets face three critical limitations: First, limited availability of HLS codes with pragma, combined with the challenges of manual labeling and expert-driven optimization of pragma coefficients, restricts dataset scalability. Second, inconsistent file structures and missing post-HLS artifacts (e.g., ADB files for CDFG reconstruction) hinder reproducibility and the applicability of these datasets for machine learning tasks. Third, existing datasets often lack comprehensive coverage of diverse application domains and HLS pragma usage.

\subsection{HLS Downstream Tasks}
HLS research focus on two main downstream tasks: QoR (quality of results) prediction and DSE (design space exploration) of HLS pragmas.
In the QoR prediction task, prior research leverages machine learning to estimate the performance and area for HLS designs early in the design cycle, without running the time-consuming high-level synthesis. Different ML techniques, such as gradient-boosted machine~\cite{7927161}, XGBoost~\cite{dai2018fast}, graph neural networks~\cite{wu2022high, sohrabizadeh2022gnn, h-gnn}, have been explored to predict post-HLS QoR at the source code level or the intermediate representation graph level. 

In the task of DSE, various configurations of HLS pragmas are evaluated and compared to find optimal design parameters with best QoR. Different DSE frameworks are proposed to improve search quality and efficiency, utilizing heuristics~\cite{zhao2017comba}, conventional methods such as genetic algorithms or simulated annealing~\cite{ferrandi2008multi, schafer2009adaptive}, or ML-based search~\cite{7927161, wu2021ironman, sohrabizadeh2022gnn, h-gnn, pouget2024automatic, pouget2024enhancing}. 
% Moreover, recent advancements further support automatic pragma insertion after the completion of DSE~\cite{pouget2024automatic, pouget2024enhancing}. 
Efficient and high-quality DSE reduces the manual effort required for tuning and optimizing HLS code, reducing the development time. 
% However, these frameworks are often built and evaluated using typical HLS benchmarks like PolyBench, which makes it difficult to assess their effectiveness on more complex designs.

\section{Dataset}

\subsection{Dataset Overview}

The data construction flow involves two key stages: collection and generation of kernels (Sec. \ref{sec:raw-data-collection}) and design space exploration for pragma insertion (Sec. \ref{sec:collection-pragma}). 
Based on this flow, we can construct a dataset that is both comprehensive and versatile, suitable for multiple HLS-related downstream tasks (Sec. \ref{sec:eval}).
As shown in Table~\ref{tab:statistics_comparison}, \textbf{ForgeHLS} significantly outperforms all existing datasets in both scale and complexity. It includes \textbf{\kernelnum{} kernels} and \textbf{\designnum{} designs}, far exceeding prior datasets in volume. More importantly, ForgeHLS offers the highest overall complexity among all datasets, with an average of \textbf{95.3 pragmas} and \textbf{718.6 tokens} per kernel. This is driven by its diverse and realistic design sources, especially in R-ForgeHLS, where Vitis-Library reaches up to \textbf{244.0 pragmas} and CHStone reaches up to \textbf{8985.8 tokens} per kernel.

Moreover, as illustrated in Figure~\ref{fig:distribution_comparison}, ForgeHLS demonstrates a broad and dense distribution across multiple post-HLS resource metrics, including LUT, DSP, FF, and BRAM\_18K utilization, as well as performance and structural attributes such as worst-case latency, best-case latency, pragma number, and code length. Unlike prior datasets that suffer from narrow, skewed, or fragmented distributions, ForgeHLS provides abundant and continuous coverage across both low and high regimes of these dimensions. This rich and diverse design space facilitates the training of more robust and generalizable models for various HLS learning tasks.

In terms of storage size, ForgeHLS occupies \textbf{2177 GB}, which is significantly larger than previous datasets. This is primarily due to the inclusion of extensive post-HLS artifacts, particularly graph-based representations such as Control Data Flow Graphs (CDFGs). These graphs are essential for enabling GNN-based learning, making ForgeHLS especially valuable for tasks like QoR prediction, HLS optimization, and performance modeling in ML research.

\begin{figure}[t]
    \centering
    \includegraphics[width=0.9\linewidth]{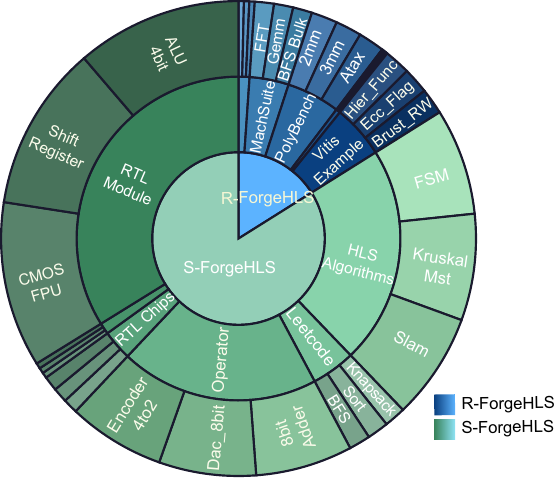}
    \caption{Breakdown of code in ForgeHLS, weighted by subset design number. The outer ring highlights 3 representative algorithms from each subset.}
    \label{fig:kernel-distribution-pie}
\end{figure}

\subsection{Collection and Generation of Kernels}
\label{sec:raw-data-collection}

\begin{table*}[t]
	\centering
	\footnotesize
	
	\begin{tabular}{@{\hspace{0em}}c @{\hspace{-0.5em}}c @{\hspace{0.7em}}c@{\hspace{0.7em}}c@{\hspace{0.7em}}c@{\hspace{0.7em}} c@{\hspace{0.7em}}c@{\hspace{0.7em}}c@{\hspace{0.7em}} c@{\hspace{0.7em}}c@{\hspace{0.7em}}c@{\hspace{0.7em}} c@{\hspace{0.7em}}c@{\hspace{0.7em}}c }
		& & \multicolumn{6}{c}{\textbf{MAPE}}& \multicolumn{6}{c}{\textbf{RMSE}}  \\
		\cmidrule(r){3-8}  \cmidrule(r){9-14}   & & \multicolumn{3}{c}{\textbf{Standard Test}} & \multicolumn{3}{c}{\textbf{Unseen Test}} & \multicolumn{3}{c}{\textbf{Standard Test}} & \multicolumn{3}{c}{\textbf{Unseen Test}} \\
		\cmidrule(r){3-5} \cmidrule(r){6-8} \cmidrule(r){9-11} \cmidrule(r){12-14} \textbf{Base Model} & \textbf{Train set} & \textbf{DSP}    & \textbf{FF}     & \textbf{LUT}    & \textbf{DSP}    & \textbf{FF}     & \textbf{LUT}    & \textbf{DSP} & \textbf{FF} & \textbf{LUT}    & \textbf{DSP}    & \textbf{FF}     & \textbf{LUT}    \\
		\midrule
				
		\multirow{3}{*}{RGCN}                                                                          & GNN Baseline       & 0.1935          & 0.2957          & 0.3197          & \textgreater 10 & \textgreater 10 & \textgreater 10 & 0.0008       & 0.0005      & 0.0008          & \textgreater 10 & \textgreater 10 & \textgreater 10 \\
		                                                                                               & S-ForgeHLS         & 0.5363          & 0.7683          & 0.4071          & 1.1822          & \textbf{1.5382}          & \textbf{0.5736}          & 0.0082       & 0.0020      & 0.0029          & \textbf{0.0710}          & 0.0323          & 0.8404          \\
		                                                                                               & ForgeHLS           & 0.6038          & 0.6188          & 0.4412          & \textbf{0.8728}          & 2.7966          & 0.6250          & 0.0181       & 0.0341      & 0.0102          & 0.1369          & \textbf{0.0303}          & \textbf{0.0395}          \\
		\midrule
				
		\multirow{3}{*}{SAGE}                                                                          & GNN Baseline       & {0.1146}          & {0.0773}          & {0.1678}          & 1.3615          & 8.1824          & 1.0354          & 0.0003       & 0.0002      & {0.0002}          & 0.0901          & 0.0489          & 0.1012          \\
		                                                                                               & S-ForgeHLS         & 0.5903          & 0.6719          & 0.2800          & 4.0069          & \textbf{1.1220}          & \textbf{0.5368}          & 0.0082       & 0.0021      & 0.0026          & 0.6072          & 0.0788          & 0.5015          \\
		                                                                                               & ForgeHLS           & 0.5448          & 0.7654          & 0.3881          & \textbf{0.9652}          & 2.3089          & 0.5441          & 0.0096       & 0.0106      & 0.0090          &\textbf{0.0790}          & \textbf{0.0204}          & \textbf{0.0417}          \\
		\bottomrule
	\end{tabular}
	\caption{Downstream Task 1. Performance comparison of GNNs for QoR prediction. \textbf{GNN Baseline}: the dataset used for GNN training in previous work~\cite{wu2022high}.  The first highest scores in unseen test are represented by \textbf{bold} font.}
	\label{tab:QoR_comparison_GNN}
\end{table*}

We adopt a dual approach to kernel collection: aggregating real-world HLS kernels from open-source repositories and generating diverse algorithmic implementations using GPT-4o. Representative algorithms are shown in Figure~\ref{fig:kernel-distribution-pie}.

For the real-world component, we aggregate kernels from previous well-established benchmarks, including:

\begin{itemize}
    \item \textbf{PolyBench}: A set of polyhedral benchmark tests designed to evaluate computational kernels.
    \item \textbf{MachSuite}: A comprehensive suite of machine learning algorithms for hardware synthesis evaluation.
    \item \textbf{CHStone}: A collection of cryptographic algorithms widely used for HLS research and FPGA design.
    \item \textbf{Rosetta}: A benchmark suite containing computer vision algorithms for performance test in hardware accelerators.
    \item \textbf{Vitis Examples}: A repository of FPGA development examples provided by Xilinx’s Vitis HLS toolchain.
    \item \textbf{Vitis Library}: A collection of performance-optimized libraries designed to accelerate various applications on AMD platforms.
\end{itemize}

Despite leveraging these diverse well-established sources, the Real-world part of ForgeHLS dataset encompassed only \textbf{405 kernels}. To extend the dataset’s breadth and enhance its applicability, we searched and collected a diverse set of HLS-compatible algorithm specifications list:

\begin{algorithm}[t]
\caption{Full DSE Explorer}
\label{alg:full_dse}
\KwIn{One C/C++ kernel $\mathcal{K}$}
\KwOut{Comprehensive set of designs $\mathcal{D}$}

$\mathit{info.tree.loop}, \mathit{info.tree.array} \leftarrow \mathit{ExtractInfo}(\mathcal{K})$ \\
$\mathcal{D} \leftarrow \varnothing$

\For{loop node $\ell$ in $\mathit{loop.tree.loop}$}{
\For{array node $\alpha$ in $\mathit{info.tree.array}$}{
$\mathit{factors}\leftarrow\mathit{GenerateFactors}(\mathit{tripcount}(\ell),\mathit{size}(\alpha))$ \\
    \For{each factor $f \in \mathit{factors}$}{
        $\mathit{pragma\_config} \leftarrow \mathit{CheckPragma}(\mathcal{K}, f)$
    }
}
}
$\mathcal{D} \leftarrow \mathit{Insert}(\mathcal{K}, \mathit{pragma\_config})$ \\

\Return $\mathcal{D}$
\end{algorithm}

\begin{algorithm}[t]
\caption{Bayesian DSE Explorer}
\label{alg:bayesian_dse}
\KwIn{One C/C++ kernel $\mathcal{K}$}
\KwOut{Comprehensive set of designs $\mathcal{D}$}

$\mathit{pragma\_config} \leftarrow \mathit{ExtractInfo}(\mathcal{K})$ \\
$\mathcal{D} \leftarrow \varnothing$ 

\For{$i = 1$ \KwTo $\mathit{N_{opt}}$}{
    $\mathit{pragma\_config.update} \leftarrow \mathit{GenerateRandomStarts}(\mathcal{K}, \mathit{pragma\_config})$ \\
    \For{$j = 1$ \KwTo $\mathit{N_{calls}}$}{
        $\mathit{DesignPoint} \leftarrow \mathit{Generate}(\mathcal{K}, \mathit{pragma\_config})$ \\
        $\mathit{RunHLS}(\mathit{DesignPoint})$ \\
        $\mathcal{D} \leftarrow \mathcal{D} \cup \mathit{DesignPoint.valid}$ \\
        $\mathit{pragma\_config.update} \leftarrow \mathit{BayesianOptimization}(\mathit{pragma\_config})$ \\
    }
}
\Return $\mathcal{D}$
\end{algorithm}

\begin{itemize}
    \item \textbf{HLS Algorithms}: Includes matrix multiplication, FFT, signal processing algorithms and deep learning kernels (e.g., LSTM, DNN layers), optimized for hardware.
    \item \textbf{Operators}: Low-level hardware components like adders, counters, and dividers.
    \item \textbf{Leetcode Algorithms}: Representative C/C++ algorithms such as binary search and merge sort, commonly used for algorithmic training.
    \item \textbf{RTL Algorithms}: A specification list of algorithms based on RTL chip, IP core, and module, as described in the RTL dataset~\cite{li2025deepcircuitx}.
\end{itemize}

Subsequently, We prompt GPT-4o with the algorithm specifications and HLS design rules, which generates the corresponding kernel code in C, devoid of pragmas. Then we run the Vitis HLS flow to validate syntactic correctness of our synthetic HLS code.

\subsection{Automated Pragma Insertion for Design Generation} 
\label{sec:collection-pragma}

After collecting kernel code, we  perform syntactic C code analysis based on Clang~\cite{clang} and LLVM~\cite{LLVM:CGO04}, and extract essential information needed for DSE, including loop trip counts, loop locations, array sizes, array definition locations, and the identification and location of the top function, as shown in Figure~\ref{fig:overview}. 

\textbf{Full DSE.} 
% With extracted code information, 
We design a tree-based explorer (Algorithm~\ref{alg:full_dse}) to systematically traverse pragma configurations for each kernel. While the traversal is near-complete, it applies HLS-specific constraints to prune invalid or redundant designs—for example, inner loops are not unrolled if the outer loop is pipelined.

However, the design space grows exponentially with the number of pragmas, making full DSE infeasible for complex kernels. For kernels with over 10 pragmas, it can produce millions of candidates, rendering exhaustive HLS synthesis computationally impractical.

\textbf{Bayesian DSE.} To mitigate the combinatorial explosion in large kernels, we propose a Bayesian design space exploration strategy (Algorithm~\ref{alg:bayesian_dse}). Instead of enumerating all pragma combinations, Bayesian DSE iteratively selects promising configurations using a surrogate model guided by prior synthesis results. Given a kernel $\mathcal{K}$, we first extract its pragma space via Clang-based static analysis. The algorithm initializes exploration with randomly sampled configurations. In each iteration, a new design is proposed, synthesized, and evaluated; the result is then used to update the Bayesian optimizer and guide subsequent pragma selection. 

Unlike conventional DSE approaches that target Pareto-optimal designs, our objective is to maximize coverage of diverse, high-quality configurations. All valid designs generated during this process are retained and included in the final ForgeHLS dataset.

We conducted over a month of DSE on a high-performance server equipped with 3 nodes, each having 128 CPU cores, totaling \textbf{over 200k CPU hours}. We are confident that our dataset will provide significant value and contribute meaningfully to the research community.

\begin{table*}[t]
	\centering
	\footnotesize
	
	\begin{tabular}{@{\hspace{0em}}c @{\hspace{0em}}c @{\hspace{0.7em}}c@{\hspace{0.7em}}c@{\hspace{0.7em}}c@{\hspace{0.7em}} c@{\hspace{0.7em}}c@{\hspace{0.7em}}c@{\hspace{0.7em}} c@{\hspace{0.7em}}c@{\hspace{0.7em}}c@{\hspace{0.7em}} c@{\hspace{0.7em}}c@{\hspace{0.7em}}c }
		& & \multicolumn{6}{c}{\textbf{MAPE}}& \multicolumn{6}{c}{\textbf{RMSE}}  \\
		\cmidrule(r){3-8}  \cmidrule(r){9-14}   & & \multicolumn{3}{c}{\textbf{Standard Test}} & \multicolumn{3}{c}{\textbf{Unseen Test}} & \multicolumn{3}{c}{\textbf{Standard Test}} & \multicolumn{3}{c}{\textbf{Unseen Test}} \\
		\cmidrule(r){3-5} \cmidrule(r){6-8} \cmidrule(r){9-11} \cmidrule(r){12-14} \textbf{Base Model} & \textbf{Train set} & \textbf{DSP}    & \textbf{FF}     & \textbf{LUT}    & \textbf{DSP}    & \textbf{FF}     & \textbf{LUT}    & \textbf{DSP} & \textbf{FF} & \textbf{LUT}    & \textbf{DSP}    & \textbf{FF}     & \textbf{LUT}    \\
		\midrule
		
		Deepseek-v3-671B                                                                               & Zero-shot          & -               & -               & -               & 3.6083          & 6.6303          & 1.4584          & -            & -           & -               & 0.2329          & 0.0253          & 0.0926          \\
		\midrule
		       
		\multirow{6}{*}{LLaMA3-7B}                                                                     & Zero-shot          & -               & -               & -               & 1.4292          & \textgreater 10 & 2.2131          & -            & -           & -               & 0.0931          & 0.0401          & 0.0527          \\
		                                                                                               & DB4HLS             & 0.7238          & 9.0168          & 6.0243          & \underline{0.7014}          & \textgreater 10 & \textgreater 10 & 0.0318       & {0.0044}      & 0.0236          & 0.0931          & 0.0401          & 0.0529          \\
		                                                                                               & HLSyn              & 8.8117          & 1.4946          & \textgreater 10 & \textgreater 10 & \textgreater 10 & \textgreater 10 & 4.2514       & 0.3345      & 0.8270          & 1.6102          & 0.0426          & 0.2521          \\
		                                                                                               & HLSDataset         & 1.2517          & 1.2488          & 0.7571          & 1.0063          & \textgreater 10 & 2.6166          & {0.0138}       & 0.0106      & {0.0158}          & \textbf{0.0918}          & \textbf{0.0398}          & \underline{0.0523}          \\
		                                                                                               & HLSFactory         & 1.2451          & 1.7471          & 1.5141          & \textgreater 10 & \textgreater 10 & \textgreater 10 & 0.0583       & 0.0144      & 0.0417          & 2.6372          & 0.9342          & 1.8684          \\
		                                                                                               & S-ForgeHLS         & 0.3135          & {0.3565}          & 0.2722          & 4.0751          & \underline{1.5098}          & \underline{0.7791}          & 0.3361       & 0.0305      & 0.1312          & \underline{0.0921}          & \underline{0.0399}          & \textbf{0.0516}          \\
		                                                                                               & ForgeHLS           & {0.2307}          & 0.6174          &{0.2604}          & \textbf{0.5511}          & \textbf{1.1123}          & \textbf{0.5970}          & 0.0821       & 0.0365      & 0.0689          & 0.0936          & 0.0410          & 0.0525          \\
		\midrule

		\multirow{6}{*}{Qwen2.5-7B}                                                                    & Zero-shot          & -               & -               & -               & \textgreater 10 & \textgreater 10 & 4.1366          & -            & -           & -               & 1.7906          & \underline{0.0385}          & 0.0536          \\
		                                                                                               & DB4HLS             & 0.8117          & 4.8090          & 3.9484          & \textbf{0.7322}          & \textgreater 10 & 9.5164          & 0.0576       & 0.0061      & 0.0253          & \textbf{0.0692}          & \textbf{0.0379}          & 0.0653          \\
		                                                                                               & HLSyn              & 9.1322          & 3.3357          & \textgreater 10 & 1.1127          & 7.5602          & 2.3977          & 2.0225       & 0.1488      & 0.4875          & 0.1351          & 0.0399          & 0.0532          \\
		                                                                                               & HLSDataset         & 1.3906          & 1.3399          & 1.0738          & 1.3438          & \textgreater 10 & 2.1044          & 0.0139       & 0.0109      & 0.0156          & 0.0917          & 0.0397          & \underline{0.0521}          \\
		
		                                                                                               & HLSFactory         & 0.4442          & 0.5438          & 0.5316          & 9.2722          & \textgreater 10 & 3.2171          & 0.0946       & 0.0145      & 0.0787          & 0.0946          & 0.0407          & 0.0533          \\
		                                                                                               & S-ForgeHLS         & 0.1230          & 0.1803          & 0.1503          & 1.0593          & \underline{1.1881}          & \textbf{0.4918}          & 0.3181       & 0.0294      & 0.1189          & \underline{0.0920}          & 0.0399          & \textbf{0.0521}          \\
		                                                                                               & ForgeHLS           & 0.4666          & 0.4937          & 0.2703          & \underline{1.0150}          & \textbf{0.8406}          & \underline{0.4960}          & 0.3394       & 0.0434      & 0.1444          & 0.1528          & 0.0399          & 0.0530          \\
		\midrule
				
		\multirow{6}{*}{Mistral-7B-v0.2}                                                               & Zero-shot          & -               & -               & -               & 8.7915          & 1.0384          & 0.9897          & -            & -           & -               & 0.2032          & 0.0399          & 0.0524          \\
		                                                                                               & DB4HLS             & \textgreater 10 & \textgreater 10 & \textgreater 10 & \textgreater 10 & \textgreater 10 & \textgreater 10 & 2.8257       & 0.0321      & 0.2167          & 4.9136          & 0.0534          & 0.1010          \\
		                                                                                               & HLSyn              & 3.1403          & 1.2435          & 1.2034          & 7.6851          & \textgreater 10 & \textgreater 10 & 2.6475       & 0.2191      & 0.5880          & 0.2581          & 0.0548          & 0.1127          \\
		                                                                                               & HLSDataset         & 3.1403          & 1.2435          & 1.2034          & 7.6851          & \textgreater 10 & \textgreater 10 & 2.6475       & 0.2191      & 0.5880          & 0.2581          & 0.0548          & 0.1127          \\
		                                                                                               & HLSFactory         & 1.0000          & 0.9758          & 2.1080          & 0.9986          & \textgreater 10 & \textgreater 10 & 0.1389       & 0.0256      & \textgreater 10 & \textbf{0.0906}          & 7.5828          & \textgreater 10 \\
		                                                                                               & S-ForgeHLS         & {0.1141} & 0.1682          & 0.1113          & \underline{0.9039}          & \textbf{0.6486}          & \textbf{0.4580}          & 0.1219       & 0.0132      & 0.0492          & 0.0919          & \underline{0.0398}          & \textbf{0.0518}          \\
		                                                                                               & ForgeHLS           & 0.5330          & 0.4575          & 0.2088          & \textbf{0.7110}          & \underline{0.9706}          & \underline{0.5881}          & 0.1183       & 0.0258      & 0.0343          & \underline{0.0917}          & \textbf{0.0398}          & \underline{0.0519}          \\
			
		\bottomrule
	\end{tabular}
	\caption{Downstream Task 1. Performance comparison of LLMs for QoR prediction. The first and second highest scores in unseen test are represented by \textbf{bold} font and \underline{underline}, respectively. Zero-shot refers to the use of base model for zero-shot inference without training, thus it do not have standard test.}
	\label{tab:QoR_comparison_LLM}
\end{table*}

\section{Evaluation}
\label{sec:eval}

\subsection{Downstream Task 1: HLS QoR Prediction}

In this downstream task, we train GNN models and LLMs on various datasets for post-HLS QoR prediction. 
% compare the performance improvements of our dataset on QoR prediction using traditional models such as GNNs and LLMs. 
Specifically, we aim to explore two key questions: 
\textbf{1)} Can GNNs capturing the complex patterns within a large-scale dataset like ours? 
\textbf{2)} Are LLMs viable for regression tasks in QoR prediction after fine-tuning?

\textbf{Standard Test and Unseen Test.} To evaluate model generalization across different train set fairly, we construct an \textbf{Unseen Test Set} by selecting 10\% of kernels from \dataset{} at the kernel level, resulting in 57 kernels along with all their corresponding designs. This ensures that no kernel in the training set overlaps with those in the Unseen Test Set, enabling a fair assessment of generalization across dataset. In parallel, similar to existing QoR prediction works, we define a \textbf{Standard Test Set} for each dataset by performing a 9:1 split of designs at the design level. Unlike the Unseen Test Set, the Standard Test Set may include designs derived from the same kernels used in training, but with different pragma configurations. It is important to note that the Unseen Test Set does not have any bias towards ForgeHLS, because the kernels using in training set 
% used for the Standard Test Set 
in \dataset{} is the remaining kernel portion not included in the Unseen Test Set.

\textbf{Experimental Setup.} We adopt the knowledge-rich approach from~\cite{wu2022high} for GNN training, where ADB files are compiled into CDFGs for input. The GNN model consists of 5 layers with a hidden dimension of 300. For LLMs, we fine-tune mainstream open-source models using LoRA~\cite{hu2021lora} with Code-QoR pairs as input, as shown in figure~\ref{fig:overview}. The LoRA rank is set to 32, and the LoRA alpha is set to 64. All experiments are conducted on 8 A800 GPUs, using LlamaFactory~\cite{zheng2024llamafactory} for fine-tuning and vLLM~\cite{kwon2023efficient} for inference.

% We evaluate prediction accuracy using two standard regression metrics: Mean Absolute Percentage Error (MAPE) and Root Mean Squared Error (RMSE). For QoR prediction, we target three primary hardware resource metrics: LUT, DSP, and FF.

\textbf{GNN Performance.} Table~\ref{tab:QoR_comparison_GNN} shows the QoR prediction performance of GNN models. On the Standard Test, the GNN Baseline outperforms \dataset{} in terms of MAPE and RMSE, but it struggles on the Unseen Test, where ForgeHLS performs better. The GNN Baseline, trained on a smaller dataset with only 18,600 kernels, overfits to the simpler patterns of the Standard Test and fails to generalize to new kernels. On the other hand, ForgeHLS performs better on the Unseen Test, suggesting that the graph-based data in ForgeHLS aids generalization. However, on the Standard Test, ForgeHLS underperforms compared to the GNN Baseline, indicating that the complexity of ForgeHLS makes it harder for the GNN to fit the seen kernels. The answer to the first question is that GNNs, while effective in certain scenarios, lack the precision to consistently predict true QoR values for a dataset as extensive as ours. Thus, this leads to the second LLMs direct prediction question.

\textbf{LLM Performance.} Table~\ref{tab:QoR_comparison_LLM} shows the QoR prediction performance of various LLMs. In zero-shot evaluations, Deepseek-v3-671B performs the best among the LLMs tested. However, after fine-tuning, open-source models significantly outperform Deepseek-v3-671B. For instance, LLaMA3-7B fine-tuned on ForgeHLS achieves a DSP MAPE of 0.5511, a notable improvement over Deepseek’s zero-shot performance of 3.6803. In the Standard Test, S-ForgeHLS exhibits the lowest MAPE (best performance), followed by ForgeHLS. On the Unseen Test, ForgeHLS consistently outperforms other datasets in terms of MAPE, while S-ForgeHLS shows slightly lower performance. Moreover, LLM predictions generally drop on the Unseen Test. For example, Qwen2.5-7B trained on ForgeHLS predicts a LUT MAPE of 0.2703 on the Standard Test, which increases to 0.4960 on the Unseen Test, though still lower than the MAPE of 0.5368 observed for GNN on the Unseen Test.
Therefore, the answer to question 2 is that LLMs, after fine-tuning, are indeed viable for regression tasks in QoR prediction, offering significant improvements and better generalization over zero-shot models.

% \textbf{LLM vs. GNN.}We evaluate the generalization ability of different trained models on the fair Unseen Test, where LLMs generally show superior performance. This can be attributed to two key factors. First, in terms of the number of learnable parameters, the GNN model in our configuration has approximately 450,000 parameters, while the LoRA fine-tuned LLMs involve over 1,000,000 learnable parameters. This significant difference in capacity indicates that LLMs have a greater ability to capture complex patterns, which is beneficial for accurate QoR prediction on the challenging Unseen Test. Second, LLMs are pretrained on vast amounts of diverse data, which enhances their generalization ability. This extensive pretraining allows LLMs to better handle unseen kernel structure, including analyzing diverse HLS code encountered in the Unseen Test, or capturing pragma information and influence effectively.

% \textbf{GNNs vs. LLMs.}.The performance drop on the Unseen Test Set emphasizes the challenges of generalizing to new, unseen kernels, especially for GNN models. In contrast, LLMs consistently demonstrate better performance on the Unseen Test Set, indicating their superior ability to generalize across diverse HLS code patterns. This advantage is likely due to their large-scale pretraining and their capacity to capture complex relationships in data.

% The QoR prediction task underscores the potential of LLMs in accurately predicting HLS performance, while also presenting new challenges for training GNN prediction models on large-scale datasets.

\subsection{Downstream Task 2: Automatic HLS Pragmas Exploration}
From the first downstream task, we can infer that LLMs possess a certain level of HLS capability. Considering pragmas, which are a crucial aspect of HLS code, evaluating the performance of LLMs in handling and optimizing pragmas becomes essential. Using existing LLMs to insert pragmas into HLS code presents several challenges: incorrect pragma insertion locations, lack understanding of hardware-specific characteristics, resulting in synthesis failure, and provide suboptimal pragma choices. 
% Additionally, they are unable to optimize pragmas effectively based on hardware resource usage to achieve a Pareto-optimal design. 

Our dataset has a large number of C kernels without pragma, and corresponding designs with pragma. It is not difficult to envision the potential for utilizing our HLS data to training an automatic pragma inserter based on LLM. 
% Out of \kernelnum{} \dataset{} kernels,
For each kernel, we define Pareto designs within our dataset as the dominant solutions in the trade-off between latency and average resource usage (ARU in Eq. \ref{eq:aru}), where no other designs can improve one of the metrics without sacrificing the other. 
% The ARU is calculated as shown in Equation~\ref{eq:aru}. 
Then we divide the designs on Pareto curves into three categories: high, medium, and low resource usage. The top one-third of the points in terms of ARU are classified as high resource usage, the middle one-third as medium, and the remaining one-third as low. Lastly, we construct Kernel-Design pairs in a format such as \noindent\lstinline|{"instruction": "optimize for low resource usage and high latency.", "input": origin code, "output": Pareto design code with HLS pragma consuming low resource usage and high latency"}|. 
% By doing so, our trained pragma inserter can generate Pareto-optimal pragma designs for varying resource usage scenarios based on the input kernel code.

We fine-tune the LLaMA3, Qwen, and Mistral models on 90\% kernels using the LoRA\cite{hu2021lora} method and evaluate their performance on the rest 10\% kernels from the same dataset.
% from \dataset{}. 
For each design generated based on the model's pragma suggestions, we run the HLS flow and measure the synthesis pass rate. To assess the quality of the generated pragma designs, we compare the predicted Pareto designs under three strategies with the ground-truth Pareto designs provided in our dataset. We use the Adjusted Distance to Reference Set (ADRS) as the evaluation metric. A lower ADRS value indicates that the generated designs are closer to the true Pareto front. The ADRS is computed as:
\begin{equation}
\fontsize{9}{9}\selectfont
\text{Average ADRS} = \frac{1}{N}  \sum_{k=1}^{N}\frac{1}{|\Gamma_k|} \sum_{\gamma \in \Gamma_k} \min_{\omega \in \Omega_k} \delta(\gamma, \omega)
\end{equation}

\begin{equation}
\fontsize{9}{9}\selectfont
\delta(\gamma, \omega) = \max\left\{ 0, \frac{l(\gamma) - l(\omega)}{l(\omega)}, \frac{r(\gamma) - r(\omega)}{r(\omega)} \right\} \times 100\%.
\end{equation}

\begin{equation} 
\fontsize{9}{9}\selectfont
\label{eq:aru} 
r(\cdot) = \text{ARU} = \frac{1}{N} \sum_{k \in {\text{BRAM}, \text{FF}, \text{LUT}, \text{DSP}}} \frac{\text{Used Resource}_k}{\text{Available Resource}_k} 
\end{equation}

where \(N\) denotes the total number of kernels, \(k\) denotes the kernel index, \(\Gamma_k\) denotes the exact Pareto-optimal set of kernel \(k\) in ForgeHLS, and \(\Omega_k\) represents the Pareto-optimal set predicted by the fine-tuned model
% in different resource usage strategy 
for kernel \(k\). The term \(\delta(\gamma, \omega)\) calculates the normalized distance between two design points. \(l(\cdot)\) refers to the cycle latency of design point, extracted from the report generated by HLS. And \(r(\cdot)\) refers to ARU of the design point. 
% The expression $\max\left\{0\right\}$ in $\delta(\gamma, \omega)$ ensures that if the proposed method generates a better pragma combination than the ground truth within our dataset, it will be considered to have a zero distance to Pareto curve, which is not penalized. 

\textbf{Increased HLS Pass Rate}. As shown in Table~\ref{tab:pragma_insertion_results}, after training, the HLS pass rate increases significantly across all LLMs. For instance, the LLaMA3-7B model trained on \dataset{} achieves a pass rate of 69\%, compared to just 19\% for the origin model. The improvement indicates that the trained LLMs are better at inserting valid pragmas. Specifically, the model becomes more adept at avoiding invalid pragma insertion locations, improper pragma formats, configuration issues, and C++ syntax errors in the HLS code. These results indicate that fine-tuned LLMs gain a better understanding of HLS code format and syntax.

\textbf{Reduced ADRS}.  Fine-tuning significantly lowers ADRS (e.g., Qwen-7B achieves 38.36\%), demonstrating LLMs' capability to generate pragma configurations closer to Pareto-optimal designs. Critically, these results derive from a one-step exploration paradigm – a simplified DSE setup. The attained low performance at ADRS reveals potential for LLMs in multi-step DSE frameworks, where iterative refinement could further optimize latency-resource trade-offs.

\textbf{Performance on Full ForgeHLS}. 
% Despite the improvements in HLS pass rates and ADRS reduction, 
The ADRS on the full ForgeHLS is 
% achieved by models fine-tuned on the full ForgeHLS dataset, 
relatively higher than that on the S-ForgeHLS (synthetic part of ForgeHLS). This is because the full dataset contains a broader spectrum of real-world kernels with diverse coding styles and optimization intents, introducing more ambiguity and complexity for learning effective pragma strategies. In contrast, S-ForgeHLS designs exhibit more regular patterns, enabling easier learning and lower ADRS.
% remain relatively high when training on the full ForgeHLS dataset, compared to only training on synthetic part of \dataset{}. 
It also highlights the richness, diversity, and complexity of our \dataset{}, which offers a broader range of HLS code patterns and kernel configurations.
% compared to traditional datasets.
% The relatively larger ADRS when training on ForgeHLS suggests that 
% More effective learning methods are needed to fully exploit the diversity and complexity of our dataset. 
This calls for the development of more advanced models or training methods to leverage the full potential of our dataset for optimizing HLS designs.

% \section{Future Work}
% \label{sec:future_work}

% In the future, we plan to expand \dataset{} by adding more kernels and incorporating additional DSE techniques to enhance dataset diversity and improve QoR prediction models. Given the dataset's complexity, we aim to explore advanced models, including enhanced GNNs and powerful LLMs, to better handle intricate design patterns and improve generalization. We will also investigate reinforcement learning and evolutionary algorithms for pragma optimization and resource allocation, further optimizing performance and design space exploration.

\begin{table}[t]
\centering

\footnotesize
\begin{tabular}{l@{\hspace{-1em}}cc}
\toprule
\textbf{Model}                 & \textbf{HLS Passed(\%)} & \textbf{Average ADRS(\%)} \\
\midrule
\multicolumn{3}{l}{\textbf{S-ForgeHLS}} \\
LLaMA3-7B-Origin                 &             77              &           131.62                                       \\
LLaMA3-7B-Finetuned              &             94              &            1.64                                     \\
Mistral-7B-Origin                &             8            &                40.00                                  \\
Mistral-7B-Finetuned             &            \textbf{97}             &               \textbf{0.62}                                   \\
\midrule
\multicolumn{3}{l}{\textbf{\dataset{}}} \\
LLaMA3-7B-Origin                 &             19              &           1554.32                                       \\
LLaMA3-7B-Finetuned              &             \textbf{69}              &            48.68                                    \\
Qwen-7B-Origin                 &             27              &           1093.96                                      \\
Qwen-7B-Finetuned                 &             54              &           \textbf{38.36}                                       \\
Mistral-7B-Origin                &             8            &                163.71                                      \\
Mistral-7B-Finetuned             &            58             &               39.62                                   \\ 
\bottomrule
\end{tabular}
\caption{Downstream Task 2. Comparison of Pragma Exploration Performance Across Different LLMs trained on Two Datasets: \textbf{\dataset{}} and \textbf{S-\dataset{}}.}
\label{tab:pragma_insertion_results}
\end{table}

\section{Conclusion}

In this paper, we introduced \textbf{ForgeHLS}, a large-scale, open-source dataset for HLS DSE and QoR prediction. By aggregating diverse real-world and synthetic HLS C++ kernels, \dataset{} surpasses existing datasets in scale, offering over \kernelnum{} kernels and \designnum{} designs, 10 times larger than prior datasets, consuming over 200k CPU hours during dataset construction. Additionally, we developed an automated pragma insertion workflow, enabling direct pragma embedding into the source code. We conducted two representative downstream tasks: QoR prediction and HLS pragma insertion. These tasks demonstrate the utility of \dataset{} in HLS optimization and highlight its potential for advancing research in the ML area.

\bibliography{aaai2026}

% \input{../checklist/ReproducibilityChecklist.tex}

\appendix
\onecolumn

\clearpage
\section{Design Space Coverage across Datasets.}. 

\begin{figure}[h!]
    \centering
    \includegraphics[]{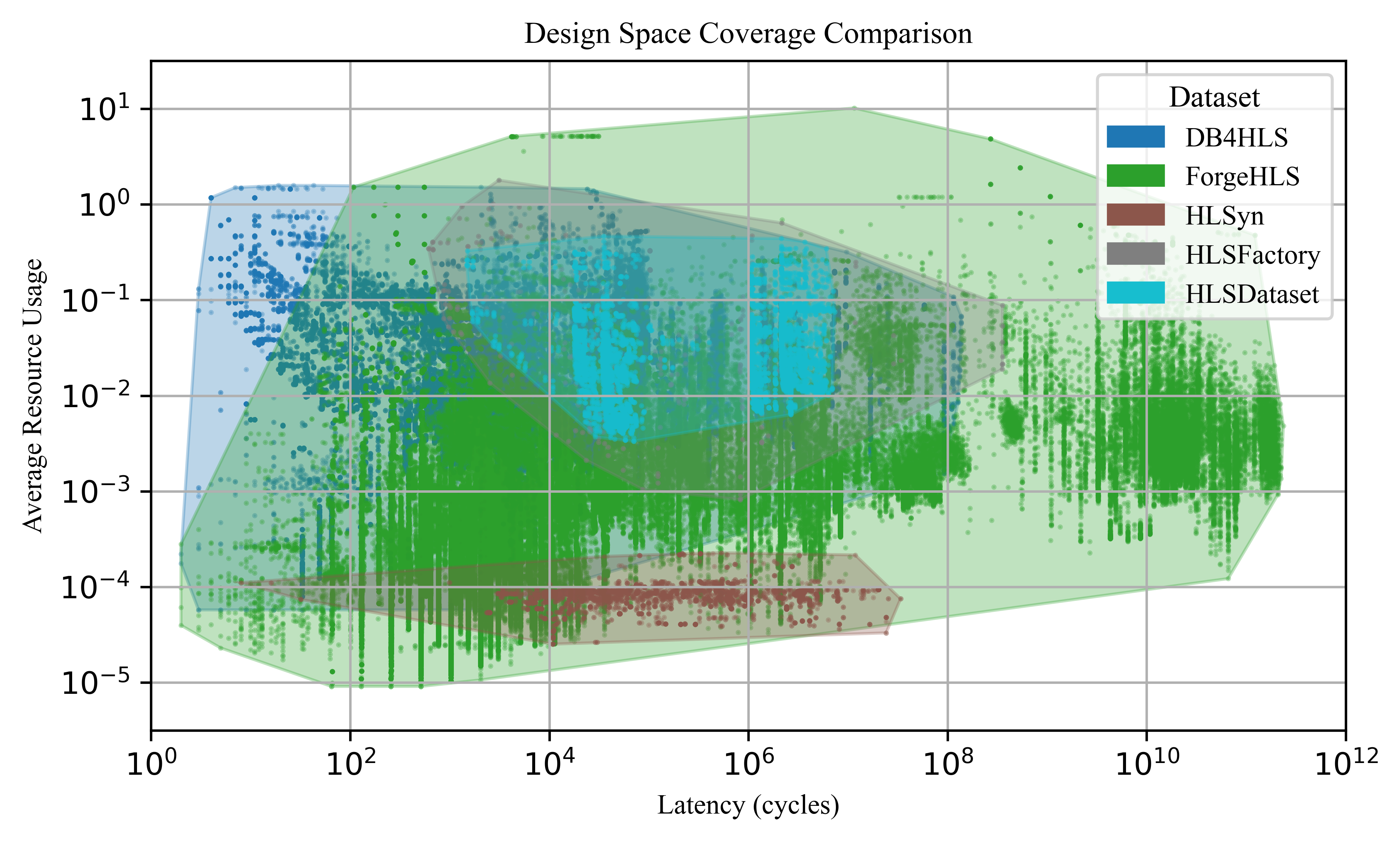}
    \caption{Design space coverage across existing datasets.}
    \label{fig:DSC_across_dataset}
\end{figure}

As shown in Fig. \ref{fig:DSC_across_dataset}, our dataset demonstrates superior coverage in terms of both latency and resource usage. Our dataset’s larger scale and more comprehensive distribution make it better suited for addressing the diverse challenges encountered in real-world applications. This broader coverage enhances the robustness of model training for tasks like QoR prediction and hardware design optimization.

We evaluate the effectiveness of our approach in expanding the design space coverage, particularly with respect to latency and resource usage, which are essential metrics for model training in downstream tasks such as QoR prediction. For latency, we use the Worst Case Latency extracted from the Vitis HLS report, and for resource usage, we compute the Average Resource Usage (ARU) across four key resource (BRAM, FF, LUT, and DSP). The ARU is calculated using the following equation \ref{eq:aru2}. $N$ represents the number of resources for which $\text{Available Resource}_k \neq 0$.

\begin{equation} 
\label{eq:aru2} 
\text{ARU} = \frac{1}{N} \sum_{k \in {\text{BRAM}, \text{FF}, \text{LUT}, \text{DSP}}} \frac{\text{Used Resource}_k}{\text{Available Resource}_k} 
\end{equation}

It is also worth noting that HLSyn’s design results in designs that tend to exhibit low resource usage. This leads to a lack of diversity in design solutions. For robust model training, it is crucial to consider not only Pareto-optimal solutions but also non-Pareto solutions. These non-Pareto solutions can also provide valuable insights into the trade-offs between resource usage and latency, contributing to the robustness of the model and its ability to generalize to a broader range of design scenarios.

In conclusion, the expanded coverage of latency and resource usage in our dataset strengthens its ability to model a wider range of hardware designs, which is crucial for advancing QoR prediction and other downstream tasks in hardware design optimization.

\clearpage
\section{Design Space Coverage of ForgeHLS}. 

\begin{figure}[h!]
    \centering
    \includegraphics[]{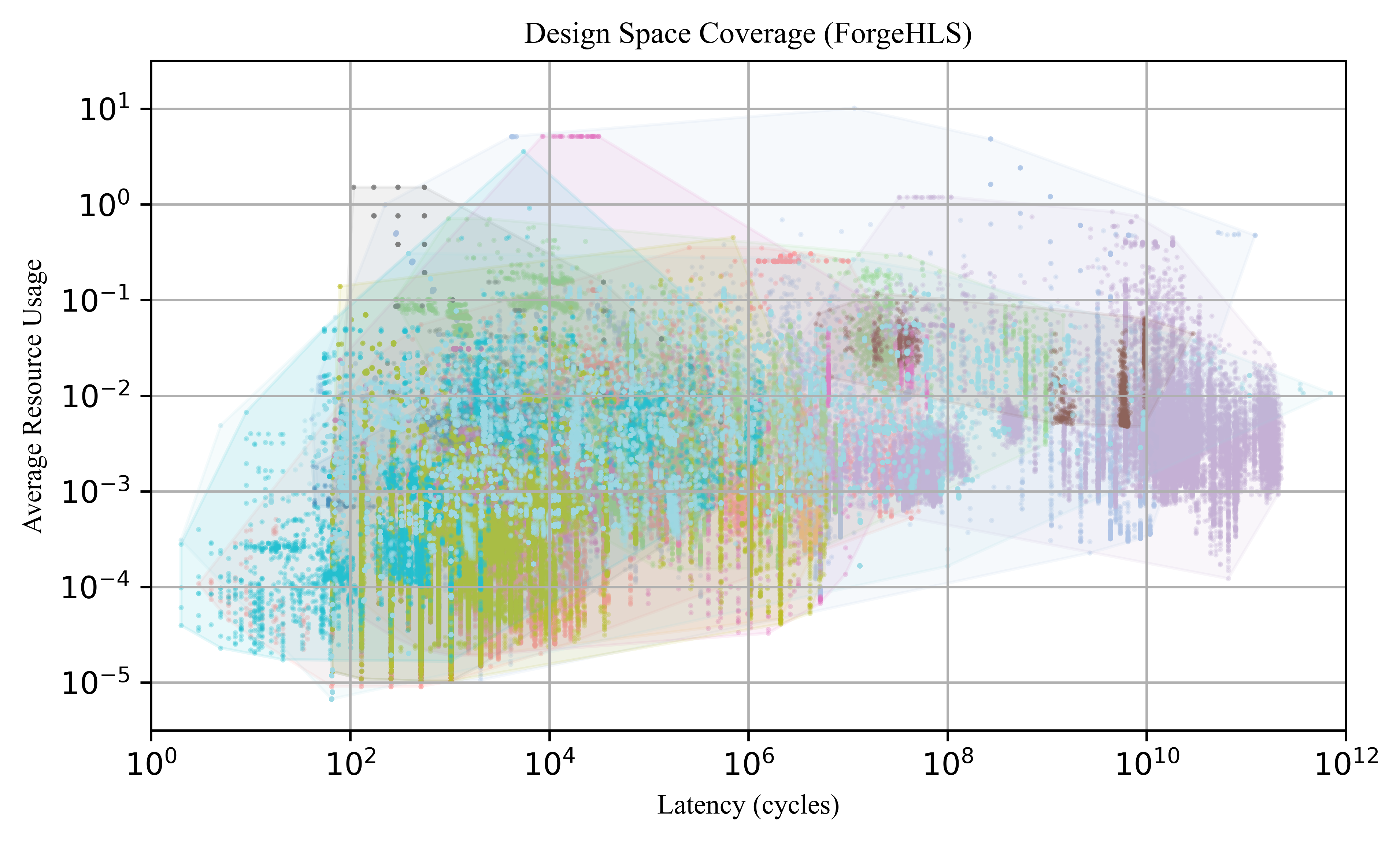}
    \includegraphics[width=0.6\linewidth]{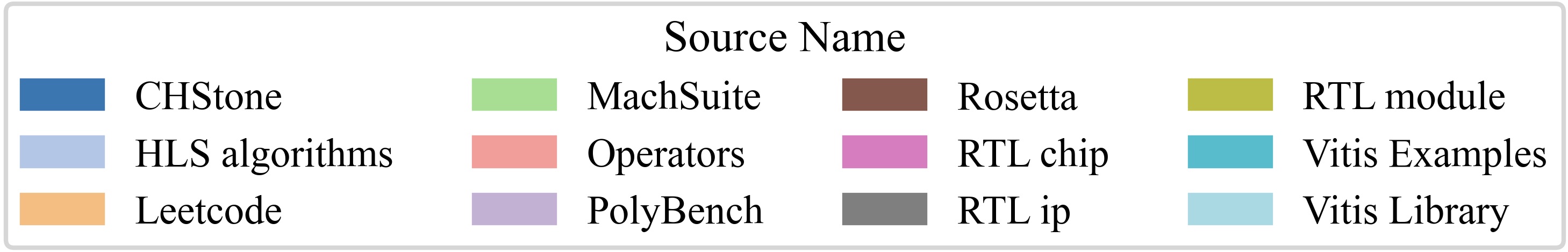}
    \caption{Design space coverage of ForgeHLS across different data sources}
    \label{fig:DSC_ForgeHLS}
\end{figure}

We have progressively expanded our dataset to achieve broader kernel coverage. Initially, we used standard HLS benchmark suites, such as CHStone, MachSuite, PolyBench, and Rosetta. We then extended the dataset by incorporating kernels from the Vitis Examples industry database, which significantly improved coverage. Furthermore, the inclusion of kernels generated by LLM Synthetic Code contributed to another substantial increase in design space coverage. As shown in Fig. \ref{fig:DSC_ForgeHLS}, these kernels, sourced from diverse origins, complement each other in expanding the design space and enriching dataset diversity.

\clearpage
\section{Detailed Distribution of \dataset{}}. 

\begin{figure}[h!]
    \centering
    \includegraphics[width=1\linewidth]{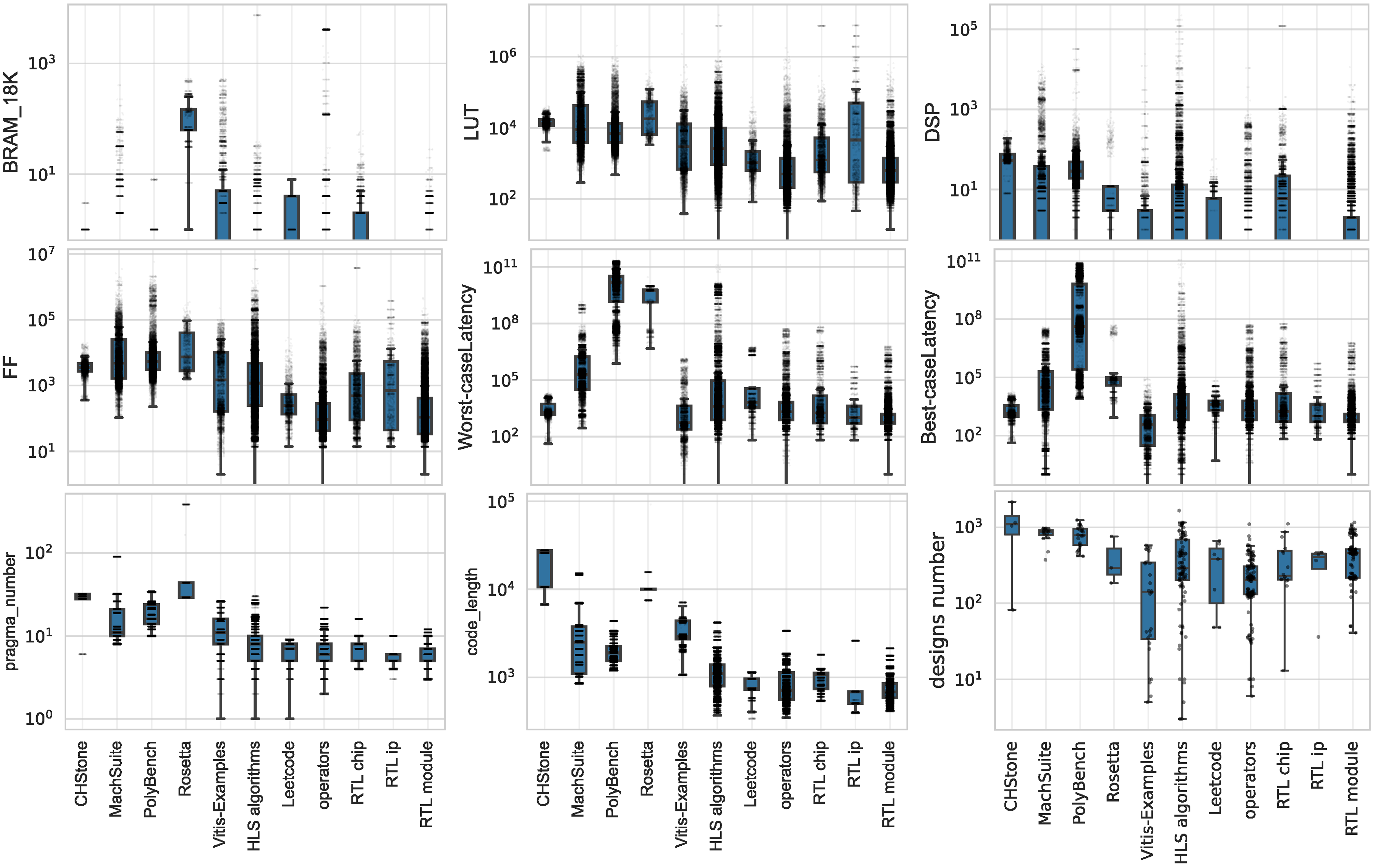}
    \caption{Detailed distribution of \dataset{} across data source. Each dot represents one individual design. Each box shows the spread of the middle 50\% of data. Metrics include \textbf{LUT} (Look-Up Table), \textbf{DSP} (Digital Signal Processor), \textbf{FF} (Flip-Flop), \textbf{BRAM\_18K} (18-Kbit Block RAM), \textbf{Worst-case Latency} in cycles, \textbf{Best-case Latency} in cycles, \textbf{Pragma Number} applied per kernel, \textbf{Code Length} per kernel by character count, \textbf{Designs Number} per kernel.}
    \label{fig:distribution_forgehls}
\end{figure}

The detailed distribution of the dataset across different sources is shown in Fig. \ref{fig:distribution_forgehls}. We present key metrics of our dataset, including the code length, latency, number of pragmas and resource usage. Each data point in these plots represents an individual design, demonstrating the spread of design characteristics across sources. This analysis shows the incremental improvement in coverage as new sources are added, particularly highlighting how LLM-generated synthetic code helps fill gaps in kernel diversity and design complexity. From this distribution, it can be observed that the synthetic code exhibits a relatively rich distribution overall. While the number of pragmas and tokens in synthetic code is slightly lower compared to real-world benchmarks, the feature distribution in terms of resource usage and latency appears comparable. This indicates that its diversity in resource and latency characteristics is sufficient for robust model training.

\clearpage
\section{ForgeHLS JSON Format Example}

\begin{lstlisting}[caption={Structured JSON example used for HLS design representation in \dataset{}},label={lst:json-example},basicstyle=\footnotesize\ttfamily]
[
    {
      "File Path": "MachSuite/aes/design_123",
      "Part": "xczu9eg-ffvb1156-2-e",
      "Avialable_BRAM_18K": 1824,
      "Avialable_LUT": 274080,
      "Avialable_DSP": 2520,
      "Avialable_FF": 548160,
      "TargetClockPeriod": 10.0,
      "EstimatedClockPeriod": 3.537,
      "Best-caseLatency": 2897,
      "Worst-caseLatency": 135246,
      "BRAM_18K": 0,
      "LUT": 3784,
      "DSP": 0,
      "FF": 874,
      "design_id": "123",
      "algo_name": "aes",
      "source_name": "MachSuite-flatten",
      "is_pareto": false,
      "is_kernel": false,
      "code_length": 6828,
      "pragma_number": 7,
      "top_function_name": "aes",
      "latency-resource-strategy": "high-latency-low-resource"
      "source_code": [
        {
          "file_name": "support.h",
          "file_content": "#include <stdlib.h>\n#include <inttypes.h>\n..."
        },
        {
          "file_name": "aes.c",
          "file_content": "void aes256_encrypt_ecb(...) { ... }\n..."
        },
        {
          "file_name": "aes.h",
          "file_content": "typedef struct { ... } aes256_context;\n..."
        }
      ]
    },
    {
    ...
    }
]
\end{lstlisting}

\clearpage
\section{Generate Synthetic Data}

We input the specifications for the algorithm description and HLS code rules into GPT-4o, which then automatically generate the corresponding code without HLS pragmas. The HLS code rules ensure correct input and output parameter usage, maintain scalability for future pragma additions, enforce strict compliance with the C++11 standard, and prohibit the use of unsupported C/C++ constructs. The unsupported C/C++ constructs include system calls, dynamic memory usage, pointer limitations, recursive functions, undefined behaviors, and virtual functions and pointers. 

\begin{lstlisting}[caption={Structured prompt for HLS code generation},label={lst:prompt-template},basicstyle=\footnotesize\ttfamily]
    prompt = f"""
Please generate a Vitis HLS code for the following algorithm:
Algorithm Name: {algorithm_name}
Requirements:
- Input and Output Parameters: The code should include Large-scale, fixed, and generated to be used as a benchmark input and output parameter types, reflecting real world usage.
- Large scale: the trip count or the size of array should be a large scale like 2^10 or less or more. Given by macro definition.
- Synthesis Readiness: The code should be a complete, standalone function or module that can be directly high level synthesized in Vitis HLS without the need for additional files or configurations. 
- Please generate a C++ program with all necessary header files correctly included, using only synthesizable constructs supported by Vitis HLS, ensuring the code can be successfully synthesized.
- Do not provide any test code. Do not provide any explanation. Only output the cppcode and the function name.
- The code should not directly include HLS pragmas
- For algorithms that may be more complex, please expand them reasonably.
- After the code is generated, please provide the function name in the comment, as shown in the example below. Format: // Top function name: function_name
- Try to avoid using while loops and use for loops instead.

Vitis HLS coding styles:
- The function and its calls must contain the entire functionality of the design.
- None of the functionality can be performed by system calls to the operating system.
- The C/C++ constructs must be of a fixed or bounded size.
- The implementation of those constructs must be unambiguous.
- Ensure the code does not use C++ features unsupported by vitis hls, such as:
    - Dynamic memory allocation (e.g., new and delete)
    - Virtual functions and polymorphism
    - Recursion
    - Complex data structures from the standard library (e.g., std::vector, std::map, std::list)
    - File I/O
    - Multithreading and concurrency (e.g., std::thread, std::mutex)
    - Pointer
    - Random number generation (e.g., std::rand, std::random)

Example:
Algorithm Name: gemm
Code Implementation:
```cpp
void gemm(int ni, int nj, int nk,
   double alpha, beta, C[ 1000 ][1100], A[ 1000 ][1200], B[ 1200 ][1100])
{{
  int i, j, k;
  for (i = 0; i < 1000; i++) {{
    for (j = 0; j < 1100; j++)
      C[i][j] *= beta;
    for (k = 0; k < 1200; k++) {{
      for (j = 0; j < 1100; j++)
        C[i][j] += alpha * A[i][k] * B[k][j];}}}}}}
// Top function name: gemm
```
"""
\end{lstlisting}

\clearpage
\section{Validate Synthetic Data}
We run a single-pass HLS flow to validate the syntactic accuracy of LLM generated code. To validate the functional accuracy, we employ a cross-verification approach leveraging GPT-4o as an assistant. Specifically, we provided GPT-4o with an algorithm description and the corresponding implementation, requesting it to generate a Python version of the same algorithm. This Python implementation served as a reference model for validation. Only when the synthetic code meets both syntactic and functional correctness criteria, we collect it in our \dataset{}. 

\begin{figure}[h]
    \centering
    \includegraphics[width=0.6\linewidth]{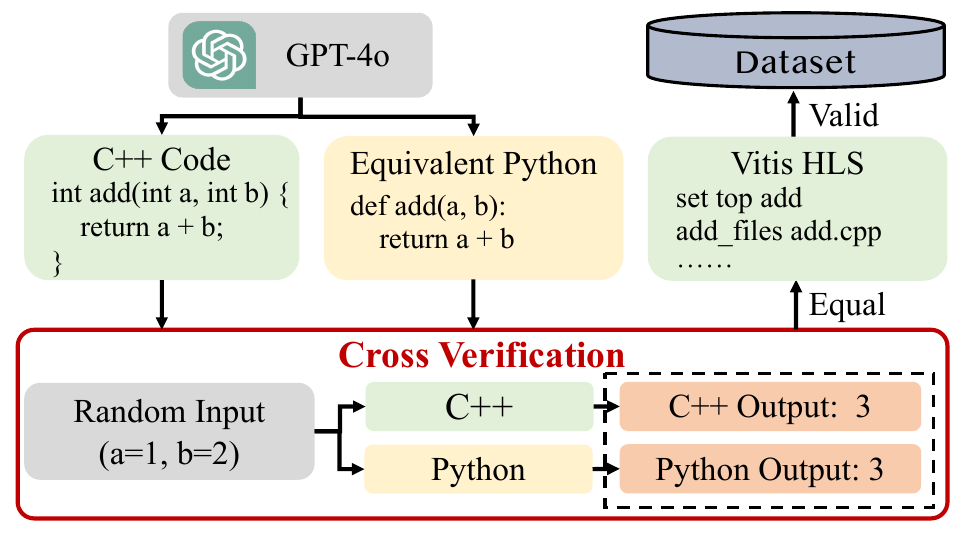}
    \caption{The structure of generator and filter.}
    \label{fig:generator-filter}
\end{figure}

\clearpage
\section{Full DSE} 

After collecting both real-world and synthetic HLS data, we perform full Design Space Exploration (DSE) and Bayesian DSE to systematically expand the design space coverage.

\begin{figure}[h!]
    \centering
    \includegraphics[width=\linewidth]{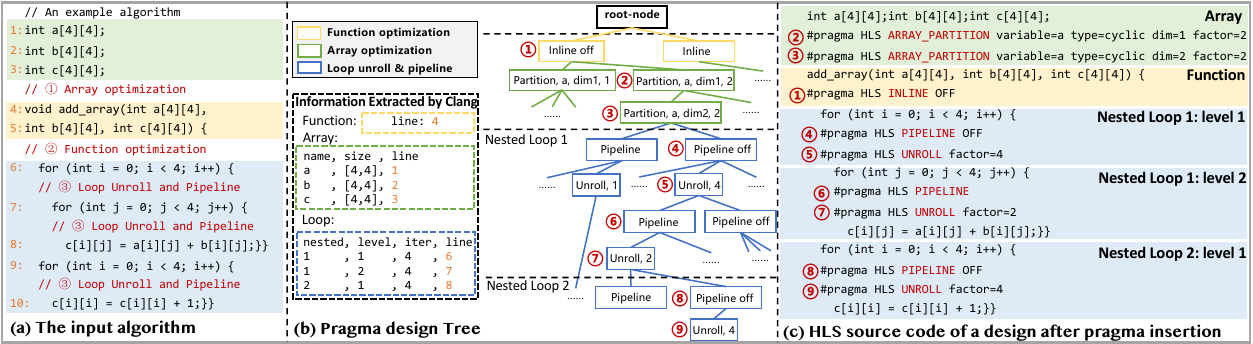}
    \caption{The process of full design space exploration: (a) showcases an input algorithm written in C++ with a nested loop structure and array manipulations; (b) illustrates the pragma design tree, detailing the hierarchical insertion with multiple optimization options; and (c) demonstrates an example of HLS code after pragma insertion, where multiple pragmas are automatically annotated into the source code, enabling diverse HLS design generation.}
    \label{fig:pragmadesign}
\end{figure}

For full DSE, we propose a pragma insertion workflow, as shown in Fig. \ref{fig:pragmadesign}. The process begins with Clang, which performs a syntactic analysis of the C++ source code to extract essential information needed for DSE. This includes loop trip counts, loop locations, array sizes and dimensions, array definition locations, and the identification and location of the top function. Such analysis ensures that the appropriate pragmas can be inserted accurately into the code.

Following this, we construct a pragma insertion tree to guide the selection of pragma factors. Specifically, we determine whether to enable or disable function inlining and pipelining on the first line inside each function. For the \texttt{array partition} pragma, we choose the target array, the dimension to apply, the partition factor, and the partition type. For loop optimizations, we decide whether to apply \texttt{loop pipeline}, whether to apply \texttt{unroll}, and if unrolling is applied, what unroll factor to choose. It is worth emphasizing that the full DSE does not exhaustively enumerate all pragma permutations. Instead, it follows a set of essential HLS design principles: 

\begin{itemize}
    \item If the outer loop in a nested loop structure is annotated with a \texttt{pipeline} pragma, the inner loop will not be assigned an \texttt{unroll} pragma.
    \item The selection of unroll and partition factors is restricted to powers of two, bounded by the loop trip count or array size.
    \item When a branch of the pragma design tree grows excessively large, pruning is applied to retain only configurations with equivalent unroll and partition factors.
\end{itemize}

By applying these rules, we eliminate invalid pragma combinations, which helps in reducing the dataset size without compromising its quality. A traversal in design tree from the root node to a leaf node represents one pragma configuration. Each path encodes a sequence of pragma decisions, and the resulting leaf node corresponds to a unique design instance. This tree-based representation enables systematic exploration of the design space, where constraints can be enforced dynamically during traversal to avoid infeasible or redundant configurations.

\clearpage
\section{Bayesian DSE} 
To automate the exploration of HLS pragma configurations, we implement a Bayesian optimization framework based on \texttt{scikit-optimize}. Given a C/C++ kernel, the pipeline extracts loop and array structures to construct a discrete design space, where each configuration consists of unroll factors, pipeline toggles, and array partition dimensions.

For each sampled configuration, the system inserts the corresponding HLS pragmas into the source code, synthesizes the design using Vitis HLS, and extracts latency and resource usage from the generated \texttt{csynth.xml}. The resource consumption is normalized with respect to the target FPGA (e.g., \texttt{xczu9eg-ffvb1156-2-e}) across BRAM, LUT, DSP, and FF. The cost function is defined as a balanced trade-off between latency and resource utilization:
\[
\text{Cost} = \sqrt{(\log_{10} \text{Latency})^2 + (\log_{10} \text{ARU})^2}
\]
where $\text{ARU}$ is the average resource usage caculated in equation~\ref{eq:aru2}.2

We use Gaussian Process-based optimization with Expected Improvement (EI) as the acquisition function. The search is initialized with 20 random configurations, followed by 40 optimization iterations. 

The implementation is submitted as supplementary material and can be invoked via:
\begin{lstlisting}[language=bash, basicstyle=\footnotesize\ttfamily]
python bayesian_DSE.py \
  --search_dir ./kernels \
  --data_path ./designs \
  --max_workers 32 \
  --bayesian_opt_number 25
\end{lstlisting}

\section{Experimental Platform Configuration}
\begin{table}[h!]
\centering

\begin{tabular}{ll}
\toprule
\multicolumn{2}{c}{\textbf{FPGA Configuration}} \\
\midrule
Vitis Version & 2023.2 \\
Clock Setting & 10 ns \\
Frequency & 100 MHz \\
FPGA Board & xcu280-fsvh2892-2L-e \\
Available BRAM\_18K & 4,032 \\
Available LUT & 1,303,680 \\
Available DSP & 9,024 \\
Available FF & 2,607,360 \\
\midrule
\multicolumn{2}{c}{\textbf{GPU Task Configuration}} \\
\midrule
GPU Model & NVIDIA A800-SXM4-80GB \\
GPU Quantity & 8 units \\
\midrule
\multicolumn{2}{c}{\textbf{CPU Task Configuration}} \\
\midrule
Compute Nodes & 3 nodes \\
Cores per Node & 128 cores \\
CPU Model & Intel Xeon Platinum 8462Y+ \\
Total Cores & 384 cores \\
\bottomrule
\end{tabular}
\caption{Experimental Platform Configuration}
\label{tab:platform_config_enhanced}
\end{table}

\end{document}